\documentclass[prl,reprint,amsmath,amssymb,twocolumn, floatfix, superscriptaddress, longbibliography]{revtex4-2}

\usepackage{graphicx}
\usepackage{dcolumn}
\usepackage{bm}
\usepackage{braket}
\usepackage{hyperref}
\usepackage{cleveref}
\usepackage[caption=false]{subfig}
\hypersetup{
    colorlinks=true,
    linkcolor=magenta,
    urlcolor=magenta,
    filecolor=magenta,   
    citecolor=blue
}
\usepackage{physics}
\usepackage{subfiles}
\crefname{figure}{Fig.}{Figures}
\crefname{equation}{Eq.}{Equations}

\begin{document}

\preprint{APS/123-QED}

\title{Erasure detection of a dual-rail qubit encoded in a double-post superconducting cavity}

\author{Akshay Koottandavida}
\thanks{These authors contributed equally to this work.\\ akshay.koottandavida@yale.edu \\ ioannis.tsioutsios@yale.edu}
\author{Ioannis Tsioutsios}
\thanks{These authors contributed equally to this work.\\ akshay.koottandavida@yale.edu \\ ioannis.tsioutsios@yale.edu}
\author{Aikaterini Kargioti}
\author{Cassady R. Smith}
\author{Vidul R. Joshi}
\author{\\ Wei Dai}
\author{James D. Teoh}
\author{Jacob C. Curtis}
\author{Luigi Frunzio}
\author{Robert J. Schoelkopf}
\author{Michel H. Devoret}
\thanks{michel.devoret@yale.edu}
\affiliation{%
 Department of Applied Physics, Yale University, New Haven, CT 06520, USA\\
 Yale Quantum Institute, Yale University, New Haven, CT 06511, USA
}%

\date{\today}

\begin{abstract}
Qubits with predominantly erasure errors present distinctive advantages for quantum error correction (QEC) and fault tolerant quantum computing. Logical qubits based on dual-rail encoding that exploit erasure detection have been recently proposed in superconducting circuit architectures, either with coupled transmons or cavities. Here, we implement a dual-rail qubit encoded in a compact, double-post superconducting cavity. Using an auxiliary transmon, we perform erasure detection on the dual-rail subspace. We characterize the behaviour of the codespace by a novel method to perform joint-Wigner tomography. This is based on modifying the cross-Kerr interaction between the cavity modes and the transmon. We measure an erasure rate of $3.981\pm0.003\,\rm (ms)^{-1}$ and a residual dephasing error rate up to $0.17\,\rm (ms)^{-1}$ within the codespace. This strong hierarchy of error rates, together with the compact and hardware-efficient nature of this novel architecture, hold promise in realising QEC schemes with enhanced thresholds and improved scaling.
\end{abstract}

\maketitle


\textit{Introduction.--}
Quantum error correction (QEC), the process of protecting quantum information from decoherence, is an essential ingredient towards fault tolerant quantum computation. QEC involves redundantly encoding logical qubits into an enlarged Hilbert space, targeting coherence that significantly exceeds that of its constituent components.  QEC codes can be roughly categorized into discrete variable codes \cite{steane_multiple-particle_1997} such as the surface code \cite{fowler_surface_2012, bonilla_ataides_xzzx_2021, krinner_realizing_2022}, or continuous variable codes such as the GKP \cite{gottesman_encoding_2001}, binomial \cite{michael_new_2016} or cat codes \cite{mirrahimi_dynamically_2014,albert_pair-cat_2019}. State-of-the-art experiments have demonstrated reduction of logical error rates by increasing code distance \cite{google_quantum_ai_suppressing_2023} and logical lifetimes longer than their physical components  \cite{ofek_extending_2016, sivak_real-time_2023, ni_beating_2023}.

Work on enhancing discrete variable codes has suggested using systems with erasure errors. These are leakage errors that can be detected in space and time. Interestingly, if erasure errors are the dominant errors in a QEC code and can be efficiently detected, the error threshold could substantially increase \cite{grassl_codes_1997}. Qubits with predominantly erasure errors, so-called ``erasure qubits", have been proposed with metastable states of neutral atoms \cite{wu_erasure_2022, ma_high-fidelity_2023, sahay_high-threshold_2023} and dual-rail qubits based on superconducting circuit quantum electrodynamics (cQED) architecture \cite{kubica_erasure_2023, teoh_dual-rail_2023}. While the dual-rail encoding has been a subject of extensive study in the quantum optics platform \cite{chuang_quantum_1996}, and have been investigated in superconducting circuits \cite{zakka-bajjani_quantum_2011, shim_semiconductor-inspired_2016,campbell_universal_2020}, it has only recently been implemented in cQED systems in the context of erasure errors \cite{levine_demonstrating_2023,chou_demonstrating_2023}.

A successful incorporation of erasure qubits into a QEC architecture requires a system that exhibits strong hierarchy of errors where the dominant errors are converted to erasures and the remaining Pauli and leakage errors exhibit orders of magnitude lower rates. To this end, superconducting cavity modes are ideal candidates to encode a dual-rail erasure qubit since they present natural noise bias, with photon loss being the dominant error mechanism \cite{teoh_dual-rail_2023, chou_demonstrating_2023}. In addition, superconducting cavities, especially those implemented in 3D geometries, exhibit longer lifetimes with lower intrinsic dephasing rates compared to transmons. Nonetheless, the non-linearity of auxiliary transmons is still a necessary ingredient for the control of cavity modes. As a result, dual-rail qubits with cavity modes suffer from additional loss channels introduced by these non-linear elements.

\begin{figure*}
\includegraphics{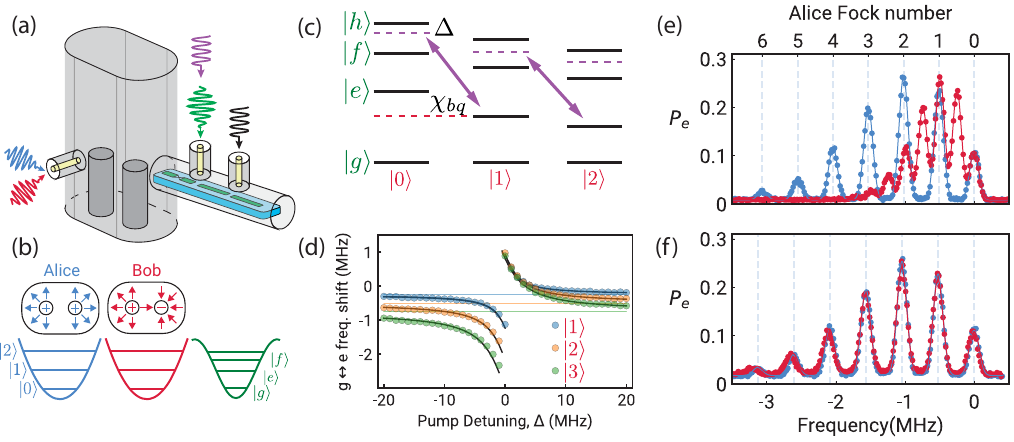}
\caption{ \textbf{(a)} Schematic of the double-post cavity, made of 5N aluminium with a transmon qubit, readout and Purcell filter fabricated on a sapphire chip. \textbf{(b)} Field distributions of the symmetric (Alice) and anti-symmetric (Bob) eigenmodes of the system, encoding the dual-rail qubit. \textbf{(c)} Energy level diagram of the combined Bob-transmon system showing dispersive shifts. The purple arrows connect the pairs of levels $\ket{n-1, h}$ and $\ket{n, e}$ being coupled via the cross-Kerr tuning drive. \textbf{(d)} Avoided crossing observed when preparing $\ket{n, e}$ states in Bob-transmon, sweeping pump detuning $\Delta$ with fixed amplitude $\Omega/2\pi = 0.5\:\mathrm{MHz}$. No drive is applied to affect Alice-transmon coupling. Solid horizontal lines are a visual aid to the bare Fock state energies and simulation results are overlaid in black lines. \textbf{(e)} Number-split peaks of the transmon when either Alice (blue) or Bob (red) is populated with a coherent state of amplitude $\alpha=1.5$, without any pump. \textbf{(f)} Bob's peaks align with that of Alice in presence of the cross-Kerr tuning pump with parameters $(\Omega/2\pi, \Delta/2\pi) = (0.5, -5.4)$ MHz.}
\label{fig1}
\end{figure*}

In this Letter, we present a dual-rail qubit implemented in a hardware efficient, 3D cavity architecture -- the symmetric and anti-symmetric eigenmodes of a double-post coaxial superconducting aluminium cavity. The highly delocalised field distributions of the modes allow for a compact architecture in which a single, dispersively coupled auxiliary transmon provides the necessary non-linearity for state preparation, erasure detection, and tomography. By modifying the dispersive interaction between the cavity modes and the transmon \cite{rosenblum_fault-tolerant_2018}, we perform joint-Wigner tomography on the two modes to characterize the erasure detection circuit. We show that our quantum non-demolition (QND) erasure detection scheme converts cavity photon losses to erasures with a false negative probability of only $0.28\%$ per gate. We demonstrate a strong hierarchy of error rates with erasures occurring at a rate of $3.981\pm0.003\,\rm (ms)^{-1}$. We measure residual Pauli errors at a rate of up to $0.170\rm (ms)^{-1}$ for the logical $\hat{X}$ and $\hat{Y}$ expectation values. Moreover, bit flip type errors occur at a negligible rate of $\sim 10^{-4}\rm (ms)^{-1}$. These results demonstrate the viability of incorporating such an erasure detection scheme in a circuit (so-called mid-circuit erasure detection) and consequently use the dual-rail as an erasure qubit in concatenation codes to enhance QEC thresholds.

\textit{Experimental System. --} \cref{fig1}(a) depicts our experiment which comprises a coaxial superconducting cavity made of high-purity (5N) aluminium with two posts of equal length. The package hosts two modes, Alice $(\hat{a})$ and Bob $(\hat{b})$, which are the harmonic oscillators used to encode the dual-rail qubit. An auxiliary transmon fabricated on sapphire chip is inserted into the package (see Supplementary Information for full system parameters). The delocalised electromagnetic field distribution of Alice and Bob (\cref{fig1}(b)) creates coupling to the transmon with similar strengths. This leads to a static dispersive interaction,

\begin{equation}
    \hat{H}/\hbar = \chi_{aq}\hat{a}^\dagger\hat{a}\ketbra{e}{e} + \chi_{bq}\hat{b}^\dagger\hat{b}\ketbra{e}{e}
\label{eq1}
\end{equation}

\noindent
with measured cross-Kerr rates $\chi_{aq}/2\pi = -0.514$MHz and $\chi_{bq}/2\pi = -0.251$MHz. However, the residual mismatch in cross-Kerr rates renders tomography in the combined Hilbert space a rather challenging task \cite{wang_schrodinger_2016, chapman_high--off-ratio_2023}.

\textit{Cross-Kerr Matching. --} To address this mismatch, we utilize a parametric process \cite{rosenblum_fault-tolerant_2018} to match the cross-Kerr rates ($\chi_{aq} = \chi_{bq}$) that allows us to perform joint-Wigner tomography on the two-mode Hilbert space. To achieve this, we leverage the four-wave mixing property of the transmon and apply a microwave drive at frequency $\omega_{\chi} = \omega_{he} - \omega_{b} + \Delta$, with amplitude $\Omega$. This process exchanges 2 photons from the transmon with 1 photon each from the drive and one of the cavities, coupling pairs of levels $\ket{n, e} \leftrightarrow \ket{n-1, h}$. Here $\ket{n}$ denotes the n-th Fock state of the relevant cavity mode, and $\ket{e}$ and $\ket{h}$ represent the first and third excited states of the transmon. The pump detuning $\Delta$ minimizes hybridization between the pairs of levels and prevents significant population exchange between them. This results in only a small energy shift of the levels from their bare values. In the $\chi << \Omega << \Delta$ regime, this can be approximated as a change in the cross-Kerr rate between the modes (\cref{fig1}(c)).

\noindent
For the purposes of this work, we choose to only tune $\chi_{bq}$  and keep $\chi_{aq}$ constant. We opted to match to the higher cross-Kerr rate of Alice in order to achieve faster gate times. This independent control of the cross-Kerr rates is possible due to the large detuning $(\approx 200$MHz) between the cavity modes. The avoided crossings in \cref{fig1}d reveal the tuning of the energy levels of the Bob-Transmon system due to the pump. We then set pump parameters $(\Omega, \Delta)$ such that the cross-Kerr rates are matched, $\chi_{bq} = \chi_{aq} = \chi$. To confirm cross-Kerr matching, we perform spectroscopic measurements on the transmon after preparing a coherent state in either cavity mode. \cref{fig1}(e) and (f) show the effect of the pump on the number split peaks \cite{gambetta_qubit-photon_2006} of the cavities, where Bob's peaks align with those of Alice. Hence we can approximate the interaction Hamiltonian as

\begin{equation}
    \hat{H}/\hbar \approx \chi \big( \hat{a}^\dagger\hat{a} + \hat{b}^\dagger\hat{b} \big)\ketbra{e}{e}
\label{eq2}
\end{equation}

\begin{figure*}
\includegraphics{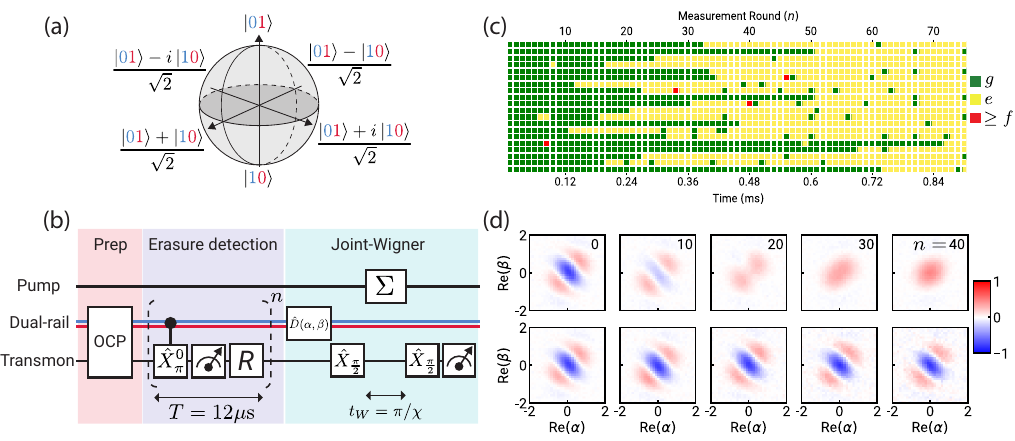}
\caption{ \textbf{(a)} Dual-rail Bloch sphere where the codewords are two-mode Fock states $\ket{+Z}_L=\ket{01}$ and $\ket{-Z}_L=\ket{10}$ (first mode is Alice and the second mode is Bob). \textbf{(b)} Erasure detection using a selective $\pi$-pulse centered at the transmon frequency $(\hat{X}^{0}_{\pi})$. After $n$ detection rounds, we perform joint-Wigner tomography on the system using the cross-Kerr matching pump. \textbf{(c)} Experimental results of erasure detection for $n=75$ rounds after preparing $\ket{+X}_L$ state. Each row is an separate run of the circuit. Ideally, state transition from $\ket{g} \rightarrow \ket{e}$ heralds an erasure for the dual-rail qubit. But readout infidelities and transmon errors causes deviation from this behaviour. \textbf{(d)} Experimental result of the full circuit shown in \textbf{(b)}. Erasure detection is performed on the $\ket{+X}_L$ state for $n$ rounds before measuring the Re$(\alpha)$ - Re$(\beta)$ joint-Wigner cut of the state. The experiment is repeated for $n=0, 10, 20, 30, 40$ rounds. The initial state clearly decays to the ground state due to cavity photon loss. Post-selecting on qubit being in $``g"$ in the trajectory data, we can reconstruct the encoded information, visually proving faithful conversion of photon loss to erasures.}
\label{fig2}
\end{figure*}

\noindent
up to $6$ Fock states. The extracted pumped cross-Kerr rates were $\chi_{aq}/2\pi = -0.521$ MHz and $\chi_{bq}/2\pi = -0.527$ MHz. We note that by increasing the pump amplitude $\Omega$ we can potentially tune higher Fock states as well.


\textit{Erasure detection. --} The dual-rail qubit is defined in the joint Hilbert space of Alice and Bob with $\ket{+Z_L} = \ket{01}$ and $\ket{-Z_L} = \ket{10}$ as the logical code words (\cref{fig2}(a)). The dominant error channel in superconducting 3D cavity modes is single-photon loss. Photon losses in either Alice or Bob destroys the logical dual-rail encoding and leaves the system in the error state $\ket{00}$. In our system, these errors occur at rates $\kappa_a = 1/T^a_1 = 4.454\pm 0.044\, \rm ({ms})^{-1} $ and $\kappa_b = 1/T^b_1 = 3.339\pm 0.018\, \rm ({ms})^{-1} $ for Alice and Bob respectively.
To detect photon losses we leverage the dispersive interaction between the cavity modes and the transmon [\cref{eq1}]. After initialising the system in the codespace, we query the transmon state using a frequency selective pulse centered at its bare frequency $(\omega_{ge})$. The transmon state flips only when the cavity modes are in $\ket{00}$ (error space). Our circuit to realise this is shown in \cref{fig2}(b). We prepare the states in the codespace via Optimal Control Pulses (OCP) \cite{khaneja_optimal_2005,heeres_implementing_2017,reinhold_reinhold-thesispdf_2019} and the transmon is initialised in its ground state $(\ket{g})$. The selective pulse to flip the transmon state is a long $(4\sigma=8\mu s)$ Gaussian pulse $(\hat{X}^0_{\pi})$. This is followed by a readout and a fast reset of the transmon state to $\ket{g}$. The entire circuit takes exactly $12 \mu$s to execute and should output readout result `$e$' if the cavity modes are in the error space and `$g$' otherwise. 

\begin{figure}
\includegraphics{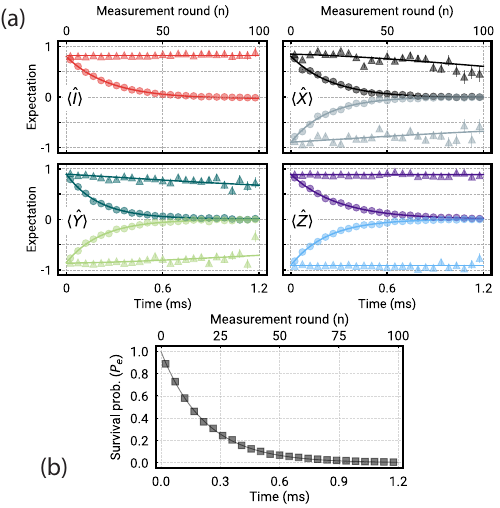}
\caption{ \textbf{(a)} Expectation values of the logical Pauli operators $\hat{I}, \hat{X}, \hat{Y}, \hat{Z}$ for the dual-rail subspace as a function of detection rounds, measured using the circuit in \cref{fig2}(b). Triangles indicate data after postselecting on no erasures from the dots. The top axes are time and the bottom is $n$ - number of detection rounds. For $\langle \hat{X} \rangle$, the experiment is performed after preparing the $\ket{+X}$(darker color at the top) and $\ket{-X}$ (lighter color at the bottom) states. The same is repeated for  $\langle \hat{Y} \rangle$ and $\langle \hat{Z} \rangle$. Postselection preserves the expectation values for far longer. The fits (shown in solid lines) for the postselected data accounts for the no-jump evolution, due to the difference in lifetimes between Alice and Bob, and residual Pauli errors within the subspace. \textbf{(b)} Survival probability, i.e., the fraction of experimental shots that survive after $n$ rounds. }
\label{fig3}
\end{figure}

\noindent
\cref{fig2}(c) displays the results of performing erasure detection for $n=75$ rounds on the $\ket{+X_L} = \frac{1}{\sqrt{2}}(\ket{01} + \ket{10})$ state. Each row in the plot represents a different experimental shot. The first row depicts a near-ideal trajectory where the transmon state remains in $\ket{g}$ until measurement round $32$. Due to a photon loss event in either cavity, the transmon frequency shifts to its bare value, causing the $\hat{X}^0_{\pi}$ gate to flip its state to $\ket{e}$. The transmon remains in this state for the remainder of the time due to the reset operation. Non-idealities arising from readout inefficiencies and transmon errors, however, cause many trajectories to deviate from the ideal behavior. To charaterize the erasure detection circuit under these errors, we feed the raw trajectories to a Hidden Markov Model (HMM) \cite{noauthor_hmmlearn_nodate}. The model learns a state transition matrix, describing the probability of transitions between codespace and errorspace at each time step, and an emission matrix, which predicts the probability of an outcome $(g/e)$ given a hidden state. From this model, we extract a false negative probability, defined as the probability of mis-assigning an erasure as being within the codespace, of $0.28\%$ per measurement. Additionally, we extract a $>99.9\%$ measure of quantum non-demolition (QND) on the cavity photon number, for our detection scheme. This means that the erasure detection itself induces minimal  backaction compared to having no detection. This is a crucial feature for incorporating such erasure checks in a surface code.


\textit{Joint tomography. --} To determine the behavior of the code space during detection, we perform direct tomography on the cavities. Since the dual-rail code is defined in the joint Hilbert space of Alice and Bob, we measure the joint Wigner function \cite{wang_schrodinger_2016}, using the circuit shown in \cref{fig2}(b). The two-mode joint Wigner function \cite{haroche_exploring_2006}, defined as,

\begin{equation}
    W(\alpha, \beta) = \frac{4}{\pi^2}\mathrm{Tr} \big[ \hat{D}(-\alpha, -\beta) \rho \hat{D}(\alpha, \beta) \hat{\Pi}\big]
\label{eq3}
\end{equation}

\noindent
is measured by displacing both Alice and Bob by $\hat{D}(\alpha, \beta) = \hat{D}(\alpha) \otimes \hat{D}(\beta)$, and measuring the joint parity operator, given by

\begin{equation}
    \hat{\Pi} = (-1)^{\hat{a}^{\dagger}\hat{a} + \hat{b}^{\dagger}\hat{b}}
\label{eq4}
\end{equation}

\noindent
A Ramsey-like sequence is employed to measure this operator. During the waiting time $(t_W)$, we establish the cross-Kerr matched interaction Hamiltonian by applying the pump with the matching parameters. For $t_W = \pi/\chi$ the unitary corresponding to \cref{eq2} maps the joint parity onto the transmon states. This circuit resembles the one employed for single-mode Wigner measurement in systems with one cavity mode \cite{vlastakis_deterministically_2013,sun_tracking_2014}, hence the necessity for matching the cross-Kerr rates. Note that the joint Wigner function is defined in a 4D space since $\alpha$ and $\beta$ are complex numbers, representing position and conjugate momentum variables of each mode. As a result, it is challenging to visualize and would require an exponential number of samples to characterize accurately.

\noindent
Instead, to efficiently characterize the system, we sparsely sample the values of $(\alpha, \beta)$. First, to visually verify the evolution of the states during erasure detection, we measure a 2D cut of the full joint Wigner space, specifically the Re$(\alpha)$ - Re$(\beta)$ cut with Im$(\alpha) =$ Im$(\beta) = 0$. In the top row of \cref{fig2}(d) we observe the evolution of the $\ket{+X_L}$ state after $n = 0, 10, 20, 30, 40$ rounds of erasure detection. As expected, the state eventually decays to the error space $\ket{00}$. After discarding trajectories where erasures where detected, we are clearly able to recover the original information (\cref{fig2}(d) bottom panel). This improvement comes at the expense of discarding more data as we track the system for longer times. Since we are only measuring a cut of the joint Wigner, we only have partial information about the system, and it is not enough to reconstruct the full density matrix. 

\noindent
Having visually confirmed that our erasure detection scheme enables faithful recovery of the encoded information, we proceed to quantify how good our detection scheme is. To achieve this, we measure the logical Pauli operators for the dual-rail subspace. This measurement is performed using the same circuit as before (\cref{fig2}(b)). By projecting the joint Wigner function in \cref{eq3} onto the dual-rail subspace, we extract the expectation values of the Pauli operators by sampling only 16 points $(\alpha_i, \beta_i)$ in phase space (see Supplementary Information) \cite{wang_schrodinger_2016,chapman_high--off-ratio_2023}.

\noindent
\cref{fig3}(a) illustrates the decay of the expectation values of the four Pauli operators, $(\hat{I}, \hat{X}, \hat{Y}, \hat{Z})$, for different states on the Bloch sphere, as a function of number of detection rounds. The expectation values decay exponentially due to the single-photon loss channels. Crucially, we observe the decay of the identity operator $(\hat{I})$, indicating that the system decays outside the code space. Averaged over the 6 cardinal states on the Bloch sphere, we obtain an erasure rate of $3.981 \pm 0.003 \, \rm ({ms})^{-1}$ or $4.8\%$ per measurement. 

\noindent
Discarding the trajectories where an erasure was detected, we extract residual Pauli error rates, within the code space. The $\hat{I}$ and the $\hat{Z}$ operator expectation values are nearly constant, and we are only able to provide an upper bound on their decay rates of  $\sim 10^{-4}\,\mathrm{(ms)}^{-1}$. The $\hat{X}$ and $\hat{Y}$ operators exhibit the expected no-jump evolution due to the difference in decay rates of Alice and Bob. In principle, it is possible to cancel this effect by designing modes with equal decay rates or by interleaving SWAP operations between the cavities such that the photon spends equal amounts of time in each mode. Fitting this deterministic no-jump evolution, we extract residual dephasing rates up to $0.170\,\mathrm{(ms)}^{-1}$ for $\langle \hat{X}\rangle$ and $\langle \hat{Y}\rangle$ ( $\approx 0.2\%$ per measurement). Note that we are discarding exponentially many trajectories as we perform erasure detection for more rounds, as shown in the survival probability plot in \cref{fig3}(b), which shows number of trajectories that survive as a function of detection rounds.


\textit{Conclusions. --}  In this work, we demonstrated erasure detection for a dual-rail qubit implemented in a superconducting double-post coaxial cavity. As a result of the compact nature of this architecture, a single auxiliary transmon is sufficient for erasure detection and control of both cavity modes. We measure an average erasure rate of $3.981\pm0.003\,\rm ({ms})^{-1}$, which corresponds to $4.8\%$ per measurement, with a false negative rate of $0.28\%$.
In addition, we developed a protocol to perform joint-Wigner tomography which relies on matching the dispersive interaction rates between cavity modes and transmon.  From 2D cuts of the joint-Wigner function space we reconstruct the encoded information given the outcomes of the erasure detection and extract the residual Pauli errors within the dual-rail code space. We observe that dephasing type errors dominate at a rate up to $0.170\,\rm ({ms})^{-1}$, $0.2\%$ per measurement, a result expected from the finite bit-flip rate of the transmon and the mismatch in the cross-Kerr rates during the erasure detection circuit. Finally, residual bit-flip and leakage errors are negligible with an upper bound of $\sim 10^{-4}\,\rm ({ms})^{-1}$, highlighting a clear hierarchy of rates where erasure dominate over Pauli errors. These results, combined with the high-fidelity beamsplitter recently demonstrated \cite{chapman_high--off-ratio_2023,lu_high-fidelity_2023}, suggest a promising pathway towards concatenating superconducting cavity based dual-rail qubits within a surface code and leverage the higher threshold and favorable scaling with code distance. Finally, it should be possible to further take advantage of the high noise-bias \cite{tuckett_tailoring_2019, bonilla_ataides_xzzx_2021, tuckett_ultrahigh_2018,puri_bias-preserving_2020},  of the Pauli errors in the dual-rail subspace when designing the surface code architecture to further improve upon the advantage of erasure detection.

We thank Shantanu Mundhada, Shraddha Singh, Shruti Puri, Alec Eickbusch, Patrick Winkel, Aniket Maiti for helpful discussions and insights throughout the project. This research was supported by the U.S. Army Research Office (ARO) under grants W911NF-18-1-0212, and W911NF-23-1-0051, by the U.S. Department of Energy, Office of Science, National Quantum Information Science Research Centers, Co-design Center for Quantum Advantage (C2QA) under contract No. DE-SC0012704, by the Air Force Office for Scientific Research (AFOSR) under grant FA9550-19-1-0399, and by the National Science Foundation (NSF) under award 2216030. The views and conclusions contained in this document are those of the authors and should not be interpreted as representing official policies, either expressed or implied, of the U.S. Government. The U.S. Government is authorized to reproduce and distribute reprints for Government purpose notwithstanding any copyright notation herein. Fabrication facilities use was supported by the Yale Institute for Nanoscience and Quantum Engineering (YINQE) and the Yale SEAS Cleanroom.

\end{document}


\preprint{APS/123-QED}

\title{%
  Supplementary Information \\
  \large ``Erasure detection of a dual-rail qubit encoded in a double-post superconducting cavity''}

\author{Akshay Koottandavida}
\thanks{These authors contributed equally to this work.\\ akshay.koottandavida@yale.edu \\ ioannis.tsioutsios@yale.edu}
\author{Ioannis Tsioutsios}
\thanks{These authors contributed equally to this work.\\ akshay.koottandavida@yale.edu \\ ioannis.tsioutsios@yale.edu}
\author{Aikaterini Kargioti}
\author{Cassady R. Smith}
\author{Vidul R. Joshi}
\author{\\ Wei Dai}
\author{James D. Teoh}
\author{Jacob C. Curtis}
\author{Luigi Frunzio}
\author{Robert J. Schoelkopf}
\author{Michel H. Devoret}
\thanks{michel.devoret@yale.edu}
\affiliation{%
 Department of Applied Physics, Yale University, New Haven, CT 06520, USA\\
 Yale Quantum Institute, Yale University, New Haven, CT 06511, USA
}

\date{\today}
\maketitle
\onecolumngrid
\tableofcontents
\clearpage

\section{Experimental setup and sample parameters\label{sec:setup}}

\begin{figure*}
\includegraphics{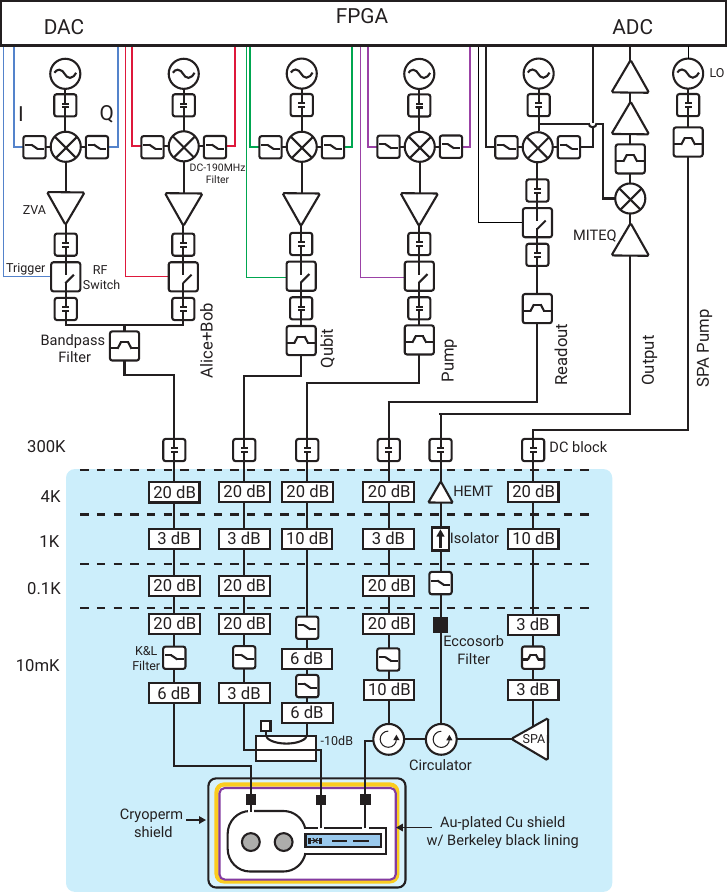}
\caption{ \textbf{Wiring diagram} Room temperature wiring for synthesizing control pulses and microwave wiring and shielding inside the dilution refrigerator.  }
\label{fig_s1}
\end{figure*}

Our experimental system consists of a coaxial stub cavity \cite{reagor_reaching_2013} with two posts, made of 5N Aluminium treated with a chemical etch to improve surface quality. The two harmonic oscillators are the two lowest frequency normal modes of the system obtained by the hybridisation of the $\lambda/4$ modes of the individual stubs. The auxiliary transmon \cite{koch_charge-insensitive_2007} made of Aluminium is fabricated on a sapphire chip along with a readout resonator and a Purcell filter \cite{purcell_resonance_1946}. The chip is inserted into a tunnel waveguide that connects to the storage cavity modes and is held in place on one side using copper clamps. The entire package is rigidly attached to a gold-plated copper bracket that is mounted on to the base plate of a dilution refrigerator. The bracket is surrounded by a gold-plated copper can, coated with a thin layer of Berkeley black on the inside to absorb high frequency photons. The outer Cryoperm shield attenuates stray magnetic fields. The top of this can is sealed with a lid with SMA feedthroughs and each seam is sealed with Indium wires. 

Control pulses for the relevant modes are synthesized via Digital-Analog Converters (DACs) from a Field Programmable Gate Array (FPGA) with a baseband of DC-250MHz \cite{reinhold_reinhold-thesispdf_2019}. These signals are up-converted to the required frequencies using IQ mixers and local oscillators (LO) which are then amplified and filtered accordingly. Fast RF switches are used to gate the signals in each line. These are then sent into different microwave lines in the dilution refrigerator each with different attenuation and filtering such that the noise temperature at relevant frequencies are around base plate temperature ($\sim$20mK). Readout signals are amplified via a quantum-limited SNAIL(Superconducting Nonlinear Asymmetric Inductive eLement) parametric amplifier(SPA) \cite{frattini_3-wave_2017,sivak_kerr-free_2019}, with a pump-port filter designed for efficient pump delivery. This amplified signal is further amplified by HEMT (High Electron Mobility Transistor) amplifier at the 4K stage and further by room temperature amplifier. This signal is then down-converted to a 50MHz signal using an IR mixer and the same LO used to up-convert the input readout pulse. After appropriate filtering and amplification, the Analog-Digital Converter(ADC) of the FPGA digitises, demodulates and integrates to obtain a readout value.

\begin{center}
\begin{table}[h]
\begin{tabular}{|c|c|l|}
\hline 
\multirow{6}{*}{\textbf{$\,\,\,$Alice$\,\,\,$}}
\rule{0pt}{3ex}
 & Frequency & $\omega_{a} =2\pi\times6.216\,\rm GHz$ \tabularnewline 
 \rule{0pt}{3ex}
 & Cross-Kerr shift & $\chi_{aq}=-2\pi\times0.514\,\rm MHz$ \tabularnewline 
 \rule{0pt}{4ex}
 & Relaxation & ${T}_{1}=224.5\pm 2.2\,\rm us$\tabularnewline
 &  & $\kappa_{a}=4.454\pm 0.044\,\rm (ms)^{-1}$\tabularnewline
 \rule{0pt}{4ex}
 & Dephasing & $T_{2}^{\,R}=452.4\pm 8.6\,\rm us$ \tabularnewline
 \rule{0pt}{4ex}
 & Thermal population & $\overline{n}_{th} = 0.0053\pm0.0002$ \tabularnewline
\hline
\multirow{6}{*}{\textbf{$\,\,\,$Bob$\,\,\,$}}
\rule{0pt}{3ex}
 & Frequency & $\omega_{b}=2\pi\times6.437\,\rm GHz$ \tabularnewline
 \rule{0pt}{3ex}
 & Cross-Kerr shift & $\chi_{bq}=-2\pi\times0.251\,\rm MHz$ \tabularnewline
 \rule{0pt}{4ex}
 & Relaxation & $T_{1}=299.4\pm 1.6\,\rm us$\tabularnewline
 &  & $\kappa_{b}=3.339\pm 0.018\,\rm (ms)^{-1}$\tabularnewline
 \rule{0pt}{4ex}
 & Dephasing & $T_{2}^{\,R}=604.6\pm 5.2\,\rm us$ \tabularnewline
 \rule{0pt}{3ex}
 & Thermal population & $\overline{n}_{th}=0.0086\pm0.0003$ \tabularnewline
\hline 
\multirow{6}{*}{\textbf{$\,\,\,$Transmon$\,\,\,$}}
 & Frequency & $\omega_{q}=2\pi\times4.681\,\rm GHz$ \tabularnewline 
 \rule{0pt}{3ex}
 & Anharmonicity & $\alpha=-2\pi\times251\,\rm MHz$ \tabularnewline
 \rule{0pt}{3ex}
 & Relaxation & $T_{1}=147.3\pm 0.5\,\rm us$ \tabularnewline
 \rule{0pt}{4ex}
 & Dephasing (Ramsey) & $T_{2R}=87.2\pm 1.3\,\rm us$ \tabularnewline
 \rule{0pt}{3ex}
 & Dephasing (Echo) & $T_{2E}=129.8\pm 1.5\,\rm us$ \tabularnewline
 \rule{0pt}{4ex}
 & Thermal population & $\overline{n}_{th}<0.01$ \tabularnewline
\hline 
\multirow{4}{*}{\textbf{$\,\,\,$Readout$\,\,\,$}} 
\rule{0pt}{3ex}
 & Frequency & $\omega_{r}=2\pi\times8.159\,\rm GHz$ \tabularnewline 
 \rule{0pt}{3ex}
 & Cross-Kerr shift & $\chi_{qr}=-2\pi\times0.432\,\rm MHz$ \tabularnewline
 \rule{0pt}{4ex}
 & Coupling strength & $\kappa_{r(c)}=3.943\,\rm (\mu s)^{-1}$\tabularnewline
 \rule{0pt}{3ex}
 & Internal loss & $\kappa_{r(i)}=1.030\,\rm (\mu s)^{-1}$\tabularnewline
\hline 
\end{tabular}
\label{tab:table1}
\caption{Measured system parameters}
\end{table}
\end{center}

\section{Cross-Kerr tuning\label{sec:chi_tuning}}
\subsection{Deriving the effective Hamiltonian\label{chi_ham}}

We begin by writing the Hamiltonian of the Alice-Bob-Transmon system with a linear drive on the transmon delivered via a capacitively coupled port. Up to the leading order in the cosine potential of the Josephson junction, the Hamiltonian is

\begin{equation}
    \hat{H}_0/\hbar = \omega_a\hat{a}^{\dagger}\hat{a}+\omega_b\hat{b}^{\dagger}\hat{b}+\omega_q\hat{q}^{\dagger}\hat{q} - g_4\bigg[\phi_a(\hat{a}+\hat{a}^{\dagger})+\phi_b(\hat{b}+\hat{b}^{\dagger})+\phi_q(\hat{q}+\hat{q}^{\dagger}) \bigg]^{4} - i\epsilon_D\cos{\omega_d t}(\hat{q} - \hat{q}^{\dagger})
\end{equation}

\noindent
where $\omega_i$ is the frequency of the $i \in \{a, b, q, d\}$ mode representing Alice, Bob, transmon and the drive, respectively, and $g_4 = E_J/4!\hbar$ is the coefficient of the $4$th order term representing the Josephson energy. First we perform a displaced frame transformation on the qubit to absorb the drive into the 4th order term. The Hamiltonion becomes 

\begin{equation}
    \hat{H}_D = \hat{D}(\alpha) \hat{H}_0 \hat{D}^{\dagger}(\alpha) + i\dot{\hat{D}}(\alpha)\hat{D}^{\dagger}(\alpha)
\end{equation}

\noindent
where $\hat{D}(\alpha) = e^{\alpha\hat{q}^{\dagger}-\alpha^{*}\hat{q}}$ and $\hat{D}(\alpha) \hat{q} \hat{D}^{\dagger}(\alpha) \rightarrow \hat{q}-\alpha$. Choosing the displacement $\alpha = \xi e^{-i\omega_d t}$ such that $\xi = \frac{i\epsilon_d}{\omega_q-\omega_d}$, ignoring terms oscillating at $2\omega_d$, we can write the Hamiltonian in the displaced frame, up to constant terms, as

\begin{equation}
    \frac{\hat{H}_D}{\hbar}  = \omega_a\hat{a}^{\dagger}\hat{a}+\omega_b\hat{b}^{\dagger}\hat{b}+\omega_q\hat{q}^{\dagger}\hat{q} - g_4\bigg[\phi_a(\hat{a}+\hat{a}^{\dagger})+\phi_b(\hat{b}+\hat{b}^{\dagger})+\phi_q(\hat{q}+\hat{q}^{\dagger} + \xi e^{-i\omega_d t} + \xi^{*} e^{i\omega_d t}) \bigg]^{4}
\end{equation}

\noindent
Next we move into a frame where the modes are rotating at their bare frequencies. We add a slight detuning to Bob from its bare detuning (adding this detuning on the transmon has equivalent resolt). The unitary for this transformation is then, $\hat{U}(t) = \exp[-i(\omega_a\hat{a}^{\dagger}\hat{a}+(\omega_b-\Delta)\hat{b}^{\dagger}\hat{b}+\omega_q\hat{q}^{\dagger}\hat{q})t]$ and we get the interaction Hamiltonian

\begin{equation}
    \frac{\hat{H}_I}{\hbar} = \Delta \hat{b}^{\dagger}\hat{b}- g_4\bigg[ \big(\phi_a\hat{a}e^{-i\omega_a t} +\phi_b\hat{b}e^{-i(\omega_b-\Delta) t} +\phi_q\hat{q}e^{-i\omega_q t} +\phi_q\xi e^{-i\omega_d t} \big) + h.c. \bigg]^{4}
\end{equation}

\noindent
where $h.c.$ represents the hermitian conjugate of all the terms in the rounded brackets preceding it. For the cross-Kerr tuning process we drive a 2-photon transition at the frequency $\omega_d = 2\omega_q - \omega_b + \Delta$. Substituting this in to the above Hamiltonian and collecting all the static terms from the expansion of the $4$th order term, with the rotating wave approximation(RWA) applied, we get

\begin{equation}
    \begin{split}
        \frac{\hat{H}_p}{\hbar} = &\,\, \Delta \hat{b}^{\dagger}\hat{b} + \frac{1}{2}\big(\Omega\hat{q}^2\hat{b}^{\dagger} + \Omega^{*}\hat{q}^{\dagger 2}\hat{b}\big) +\\
      &\,\, \Delta^a_s \hat{a}^{\dagger}\hat{a} + \Delta^b_s \hat{b}^{\dagger}\hat{b} + \Delta^q_s \hat{q}^{\dagger}\hat{q} +\\
      &\,\, \chi_{aq} \hat{a}^{\dagger}\hat{a}\hat{q}^{\dagger}\hat{q} + \chi_{bq} \hat{b}^{\dagger}\hat{b}\hat{q}^{\dagger}\hat{q} + \chi_{ab} \hat{a}^{\dagger}\hat{a}\hat{b}^{\dagger}\hat{b} +\\
      &\,\, \frac{K_{aa}}{2} \hat{a}^{\dagger 2}\hat{a}^2 + \frac{K_{bb}}{2} \hat{b}^{\dagger 2}\hat{b}^2 + \frac{K_{qq}}{2} \hat{q}^{\dagger 2}\hat{q}^2 
    \end{split}
\end{equation}

\noindent
where, $\Delta^{i}_s = \frac{1}{2}K_{ii}|\xi|^2$ is the Stark-shift and $K_{ii} = -\frac{1}{2}E_J\phi^4_i$ is the self-Kerr of the $i$-th mode and $\chi_{ij} = -E_J\phi^2_i\phi^2_j$ is the cross-Kerr between modes $i$ and $j$. In the first line in the above equation, $\Omega = \xi^{*} \sqrt{2\chi_{bq}K_{qq}}$ is the interaction rate of the cross-Kerr tuning process, detuned by $\Delta$. Going into the interaction frame with respect to the transmon self-Kerr $U(t) = \exp(-i\frac{K_{qq}}{2}t\hat{q}^{\dagger 2}\hat{q}^2)$, we get

\begin{equation}
    \begin{split}
        \frac{\hat{H}_p}{\hbar} = &\,\,\Delta \hat{b}^{\dagger}\hat{b} + \Big[\frac{\Omega}{2}\hat{b}^{\dagger}\big(\sqrt{2}\ketbra{g}{f}e^{-iK_{qq}t} + \sqrt{6}\ketbra{e}{h}e^{-3iK_{qq}t} + .. \big)\Big] + h.c. + \\
      &\,\, \chi_{aq} \hat{a}^{\dagger}\hat{a}\hat{q}^{\dagger}\hat{q} + \chi_{bq} \hat{b}^{\dagger}\hat{b}\hat{q}^{\dagger}\hat{q} + \chi_{ab} \hat{a}^{\dagger}\hat{a}\hat{b}^{\dagger}\hat{b} + \frac{K_{aa}}{2} \hat{a}^{\dagger 2}\hat{a}^2 + \frac{K_{bb}}{2} \hat{b}^{\dagger 2}\hat{b}^2
    \end{split}
    \label{s_eq_chi_1}
\end{equation}

\noindent
By choosing $\Delta \rightarrow \Delta - 3K_{qq}$ and moving into the appropriate rotating frame, we can selectively pick out the $\ket{e}\leftrightarrow\ket{h}$ transition to address and ignore the other pumped terms under the RWA($\Delta << K_{qq} << \omega_q$). Hence, we get,

\begin{equation}
    \frac{\hat{H}_{\chi}}{\hbar} \approx (\Delta-3K_{qq}) \hat{b}^{\dagger}\hat{b} + \bigg( \Omega\ketbra{e}{h}\hat{b}^{\dagger} + \Omega^*\ketbra{h}{e}\hat{b}\bigg) + \chi_{aq} \hat{a}^{\dagger}\hat{a}\hat{q}^{\dagger}\hat{q} + \chi_{bq} \hat{b}^{\dagger}\hat{b}\hat{q}^{\dagger}\hat{q}
    \label{s_eq_chi_ham}
\end{equation}

\noindent
where we've redefined the drive amplitude to $\Omega \rightarrow \sqrt{\frac{3}{2}}\Omega$ and dropped the negligible terms like the self-Kerr rates of the cavities and the cross-Kerr between them. Note that the pump frequency is $\omega_d = 2\omega_q - \omega_b - 3K_{qq} + \Delta$. By diagonalizing the above pumped Hamiltonian, we obtain the behavior of the combined Fock states, as seen from the fits in Fig1.(c) of the main text. 

To get an intuition of how the above Hamiltonian modifies the cross-Kerr, let us consider the transmon coupled to just Bob, expand in the Fock basis and select only the term that couples $\ket{h0}\leftrightarrow\ket{e1}$. Here we write a simplified form of the above Hamiltonian for this term, 

\begin{equation}
    \frac{\hat{H}_p}{\hbar} = \Delta \ketbra{e1}{e1} + \Omega\ketbra{e1}{h0} + \Omega^*\ketbra{h0}{e1}
\end{equation}

\noindent
This describes two levels coupled to each other and we can diagonalize easily to find the eigenvalues and eigenstate. The eigenvalues are $\lambda_{\pm} = -\frac{\Delta}{2}\pm\frac{1}{2}\sqrt{\Delta^2 + 4|\Omega|^2}$. For the limit of $\Omega << \Delta$, the lowest eigenvalue can be approximated to the lowest order as $\lambda_{+} \approx |\Omega|^2/\Delta$ and the corresponding eigenstate is $\ket{\lambda_+}=\ket{\Tilde{e1}} \approx \ket{e1} + \frac{\Omega}{\Delta}\ket{h0}$. For large detunings, the hybridisation is weak and if the $\ket{h}$ level of the transmon is never occupied, then we can approximately rewrite the Hamiltonian as 

\begin{equation}
    \frac{\hat{H}_p}{\hbar} \approx \frac{|\Omega|^2}{\Delta} \Big( \ketbra{\Tilde{e1}}{\Tilde{e1}} - \ketbra{\Tilde{h0}}{\Tilde{h0}} \Big)
\end{equation}

\noindent
This corresponds to modified  bare energy of the state. If we now perform the same for higher Fock states it will result in modified cross-Kerr due to the drive.

\subsection{Characterizing the cross-Kerr tuning process\label{chi_char}}

\begin{figure*}
\includegraphics{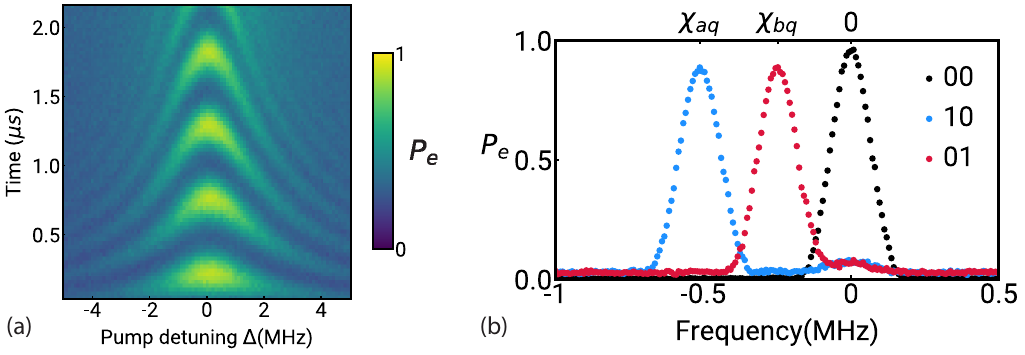}
\caption{ \textbf{a,} Chevron plot obtained after initialising the transmon in $\ket{e}$ and Bob cavity in $\ket{1}$ and sweeping time and frequency of the cross-Kerr tuning pump. Fit Rabi rate $\Omega = 2\pi \times 0.98$ MHz. Transmon spectroscopy after preparing Alice and Bob in states $\ket{00}$, $\ket{01}$ and $\ket{10}$. The Gaussian pulse used for spectroscopy is the same one used  for erasure detection. The frequency axis referenced with respect to the bare transmon $g\rightarrow e$ frequency.}
\label{fig_s2}
\end{figure*}

To characterize the cross-Kerr tuning, we prepare either of the cavities in the $\ket{1}$ state and the transmon in $\ket{e}$. By applying the cross-Kerr pump with varying frequency and time, we measure the probability of the transmon to be in $\ket{e}$. This reveals the chevron between the levels $\ket{e1}$ and $\ket{h0}$ with a Rabi rate $\Omega_{h0e1}$. We use this method to calibrate the strength and the stark-shifted centre frequency of the process, for either cavity modes using \cref{s_eq_chi_1}. \cref{fig_s2}a shows the chevron pattern of such an oscillation perform on the Bob cavity.

\subsection{Gaussian pulse for erasure detection}

\cref{fig_s2}b shows transmon spectroscopy results obtained after preparing the cavities in $\ket{00}, \ket{01}$ and $\ket{10}$ states. The spectroscopy is performed using a Gaussian pulse with $\sigma=2\mu\mathrm{s}$, with a total duration of $4\sigma=8\mu\mathrm{s}$. This is the same pulse used in the erasure detection circuit.

\section{Dual-rail qubit\label{sec:dualrail}}
\subsection{Encoding and error-channels\label{sec:error}}

We define the dual-rail qubit in the joint-Hilbert space of Alice and Bob. The logical codewords are $\ket{+Z_L}=\ket{01}$ and $\ket{-Z_L}=\ket{10}$. Here the first mode in the ket is Alice and the second is Bob. The full error set for the code due to error channels of the cavities is $E = \{\hat{I}, \hat{a}, \hat{b}, \hat{a}^{\dagger}, \hat{b}^{\dagger}, \hat{n}_a, \hat{n}_b, e^{-\frac{1}{2}(\kappa_a\hat{n}_a + \kappa_b\hat{n}_b)t}\}$. These represent the identity, cavity photon loss, cavity photon gain, cavity dephasing and the no-jump error channels respectively. The $i$-th error channel $\hat{E}_i$ will act on an arbitrary state in the logical codespace, $\ket{\psi} = V\ket{01} + W\ket{10}$, and take the system into the error space as follows,

\begin{equation}
    \ket{E_i} = \frac{\hat{E}_i\ket{\psi}}{\sqrt{\bra{\psi}\hat{E}^{\dagger}_i\hat{E}_i\ket{\psi}}}
\end{equation}

where $E_i$ is the $i$-th error channel in the set $E$. From this, it becomes apparent that the error state for the photon loss channel, in either Alice or Bob, is the ground state of the system : $\ket{E_1} = \ket{E_2} = \ket{00}$. For the photon gain channels, we have

\begin{equation}
    \ket{E_3} = \frac{V}{\sqrt{1+|W|^2}}\ket{11} + \frac{\sqrt{2}W}{\sqrt{1+|W|^2}}\ket{20}     
\end{equation}

\begin{equation}
    \ket{E_4} = \frac{\sqrt{2}V}{\sqrt{1+|V|^2}}\ket{02} + \frac{W}{\sqrt{1+|V|^2}}\ket{11}  
\end{equation}

Note that for both the photon gain channels the total excitation in the cavities is 2. Dephasing error channel of the individual cavities will lead to dephasing within the dualrail codespace. Finally, there will be a backaction on the codespace due to the no-jump evolution of the states which will look like

\begin{equation}
    \ket{E_7} = \frac{V\ket{01} + We^{-\frac{1}{2} \Delta\kappa t}\ket{10}}{\sqrt{|V|^2 + |W|^2e^{-\Delta\kappa t}}}
\end{equation}

\noindent
where $ \Delta\kappa =\kappa_a-\kappa_b$ is the difference in loss rates between the two modes. This no-jump evolution causes the state to be polarised to the longer lived mode, distorting the encoded information.

\subsection{Erasure detection and trajectories\label{sec:trajs}}

\subsection{Hidden Markov decoding\label{sec:hmm}}

\begin{figure}
\includegraphics{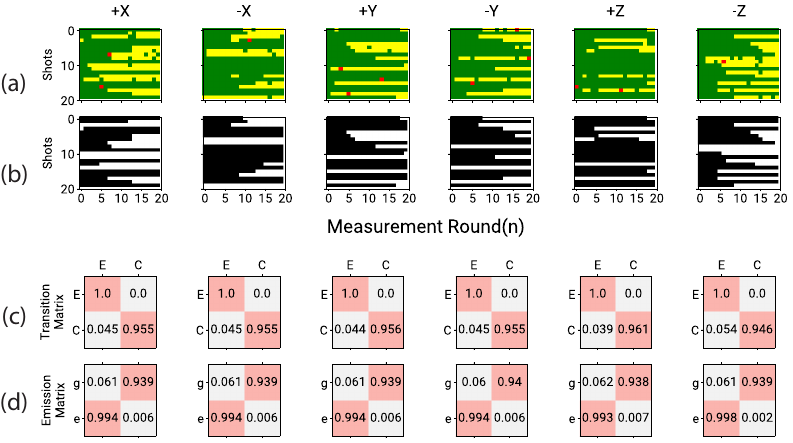}
\caption{ \textbf{a,} Raw trajectories measured after preparing the 6 cardinal states in the dual-rail subspace. Here we only show 20 samples trajectories with 20 erasure detection rounds in each trajectory. Each pixel represents a detection round with $``g"$ as green, $``e"$ as yellow and red represents states higher than $``e"$. \textbf{b,} Corresponding decoded trajectories as predicted by the Hidden Markov Model (HMM). The model is trained for each states with $10^4$ trajectories and more than 150 erasure detection rounds per trajectory. Black squares represents the system is within the dual-rail codespace and white squares represents the error space. \textbf{c,} Learned transition matrix for each of the 6 states. The off-diagonal elements represents the probability of the system to transition between codespace(C) and errorsapce(E). \textbf{d,} Learned emission matrix corresponding to the measurement outcomes $``g"$ and $``e"$. }
\label{fig_s3}
\end{figure}

To predict cavity photon loss and the most likely state of the system, we train a Categorical Hidden Markov Model (HMM) with the experimentally obtained trajectories. The HMM assumes 2 hidden states, Codespace $(C)$ and Errorspace $(E)$ and 2 measurement outcomes $``g"$ and $``e"$. The transition matrix element $t_{ij}$ determines the probability of the system to make a transition from state $i\rightarrow j$ in the current time step. Similarly, the emission matrix element $e_{mn}$ is the probability of observing the measurement outcome $m$ given that the system is in state $n$ at the current time step. In our experiment, we measure $10^{4}$ trajectories for each of the 6 cardinal states in the dual-rail Bloch sphere and train an HMM for each state to learn the probabilities. Each trajectory is 167 measurement rounds long. \cref{fig_s3}a shows a sample of these trajectories for the 6 cardinal states while \cref{fig_s3}b shows the decoded trajectories as predicted by the HMM. \cref{fig_s3}c and \cref{fig_s3}d depicts the transition and emission matrices obtained after the training on the raw trajectories. Since each erasure detection round is $12\mu s$ long, we can convert these probabilities in to rates.

\begin{table}[h]
    \centering
    \begin{tabular}{ccc}
    \hline 
    \rule{0pt}{3ex}
 State&\,\, Erasure prob.&Erasure rate\\
 \hline 
         $\ket{+Z}$&\,\,  $0.039$ per gate& $3.25 (\mathrm{ms})^{-1}$\\
         \rule{0pt}{3ex}
         $\ket{-Z}$&\,\,  $0.054$ per gate& $4.5(\mathrm{ms})^{-1}$\\
         \rule{0pt}{3ex}
         Equator states&\,\,  $0.045$ per gate& $3.75 (\mathrm{ms})^{-1}$\\
         \rule{0pt}{3ex}
    \end{tabular}
    \caption{Error probabilities calculated from the HMM model}
    \label{tab:my_label}
\end{table}

We expect the erasure rates to be $\kappa_a$ and $\kappa_b$ for the $\ket{\pm Z}$ states and $\frac{1}{2}(\kappa_a+\kappa_b)$ for the states on the equator. The HMM predicted values matches well with the rates extracted independently(from Table 1). The HMM also allows us to extract false positive and false negative probabilities. 

\begin{equation}
    \begin{split}
        \mathrm{False \, negative\, prob.} = &\,\, P(C\rightarrow E) \times P(g|E) = t_{CE}\times e_{gE} = 0.0028 \mathrm{\,per\,gate} \\
        \mathrm{False \, positive\, prob.} = &\,\, P(C\rightarrow C) \times P(e|C) = t_{CC}\times e_{eC} = 0.006 \mathrm{\,per\,gate}\\
    \end{split}
\end{equation}

The above values are averaged over all 6 cardinal states. 

\section{Joint-Wigner tomography\label{sec:joint_wig}}



\subsection{Logical Pauli measurements\label{sec:log_pauli}}

To extract the expectation values of the logical Pauli operators of the dual-rail subspace, we sample specific points in the joint-Wigner space of the cavities \cite{wang_schrodinger_2016, chapman_high--off-ratio_2023}. To see this, we first define describe how to measure the expectation values of the logical Pauli operators for the Fock qubit using single-mode Wigner. We will then extend this method to measure the logical Pauli operators of the dual-rail qubit using Joint-Wigner.

\subsubsection{Fock qubit}

\begin{figure}
\includegraphics{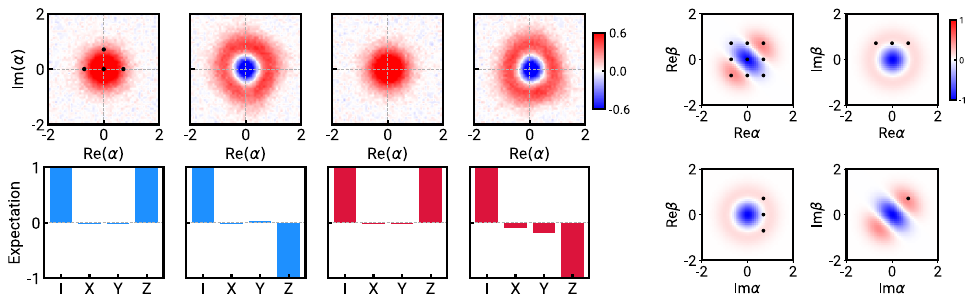}
\caption{ \textbf{a,} Measured single-mode Wigner functions for Alice and Bob after initialising in Fock states $\ket{0}$ and $\ket{1}$. The first plot shows the 4 points to be measured(black dots) in order to extract the expectation values of the Pauli operators within the $0-1$ Fock qubit. The bottom panel shows the expectation values of the Pauli operators using these 4 measurements for both Alice and Bob initialised in the same states(with $\ket{+Z}=\ket{0}, \ket{-Z}=\ket{1}$). We obtain the expected values up to SPAM errors. \textbf{b,} Simulated 2D cuts of joint-Wigner function for the $\ket{+X}_L$ state in the dual-rail subspace. In each of the plot, the other 2 values are set to 0. The black dots represents the 16 values in the joint phase space of Alice and Bob to be measured to extract the 16 2-qubit Pauli operator expectation values.}
\label{fig_s4}
\end{figure}

The single mode Wigner for a bosonic mode is defined as 

\begin{equation}
    W(\alpha) = \frac{2}{\pi}\mathrm{Tr}\big[ \rho \hat{P}_A(\alpha) \big]
\end{equation}

where $\hat{P}_A(\alpha) = \hat{D}(-\alpha)\hat{\Pi}_A \hat{D}(\alpha)$ is the displaced parity operator with $\hat{\Pi}_A = (-1)^{\hat{a}^{\dagger}\hat{a}}$ being the parity operator for the mode. The logical codewords for the Fock qubit are $\ket{+Z_L} = \ket{0}$ and $\ket{-Z_L} = \ket{1}$. We then proceed to project the displaced parity onto the fock qubit subspace using the identity projector $\hat{I} = \ketbra{0} + \ketbra{1}$. We get

\begin{gather}
 \hat{I} \hat{P}_A(\alpha)\hat{I}
=
  \begin{bmatrix}
   \bra{0}\hat{P}_A\ket{0} &
   \bra{0}\hat{P}_A\ket{1} \\
   \bra{1}\hat{P}_A\ket{0} &
   \bra{1}\hat{P}_A\ket{1} 
   \end{bmatrix}
=
   \begin{bmatrix}
   \bra{\alpha}\hat{\Pi}_A\ket{\alpha} &
   \bra{\alpha}\hat{\Pi}_A\hat{D}(\alpha)\ket{1} \\
   \bra{1}\hat{D}(-\alpha)\hat{\Pi}_A\ket{\alpha} &
   \bra{1}\hat{D}(-\alpha)\hat{\Pi}_A \hat{D}(\alpha)\ket{1} 
   \end{bmatrix}
\end{gather}

Using Laguerre polynomials and the identity $(-1)^{\hat{a}^{\dagger}\hat{a}}\ket{\alpha} = \ket{-\alpha}$, we simplify the above equation to

\begin{gather}
 \hat{I}_L \hat{P}_A(\alpha)\hat{I}_L
=
  \begin{bmatrix}
   \bra{\alpha}\ket{-\alpha} &
   \bra{-2\alpha}\ket{1} \\
   \bra{1}\ket{-2\alpha} &
   \bra{1}\hat{P}(\alpha)\ket{1} 
   \end{bmatrix}
=
e^{-2|\alpha|^2}
   \begin{bmatrix}
   1 &
   2\alpha^* \\
   2\alpha &
   4|\alpha|^2 - 1
   \end{bmatrix}
\end{gather}

here we used $\braket{-\alpha}{\alpha} = e^{-2|\alpha|^2}$ and $\bra{1}\hat{P}(\alpha)\ket{1} = e^{-2|\alpha|^2}(4|\alpha|^2 - 1)$. We then equate the above equation to the 4 Pauli operators $\{\hat{I}, \hat{X}, \hat{Y}, \hat{Z}\}$ and solving for $\alpha$, we get $\alpha_i = \{0, \frac{1}{\sqrt{2}}, -\frac{1}{\sqrt{2}}, \frac{i}{\sqrt{2}}\}$. This tells us that if we restrict ourselves to the Fock qubit subspace then we only need to measure these four points in phase space to reconstruct the full density matrix. \cref{fig_s4}a, shows the pictorially the four points in phase space and also provides intuition to why we expect it. Now, using these displacements, we can invert the above equation to get the Pauli operators as

\begin{equation}
    \begin{split}
        \hat{I} = &\,\,\frac{e}{2}\Big[ \hat{P}(\frac{1}{\sqrt{2}}) 
        + \hat{P}(-\frac{1}{\sqrt{2}}) \Big] \\
        \hat{X} = &\,\,\frac{e}{2\sqrt{2}}\Big[ \hat{P}(\frac{1}{\sqrt{2}}) - \hat{P}(-\frac{1}{\sqrt{2}}) \Big] \\
        \hat{Y} = &\,\,\frac{e}{2\sqrt{2}}\Big[ 2\hat{P}(\frac{i}{\sqrt{2}}) - \hat{P}(\frac{1}{\sqrt{2}}) - \hat{P}(-\frac{1}{\sqrt{2}}) \Big] \\
        \hat{Z} = &\,\,\hat{P}(0) \\
    \end{split}
\end{equation}

\subsubsection{Dual-rail qubit}

\begin{figure}
\includegraphics{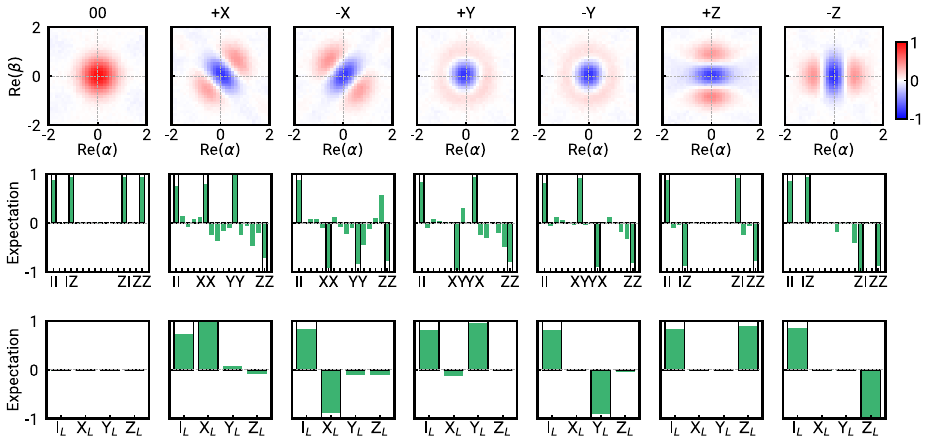}
\caption{ \textbf{a,} Experimentally measured 2d joint-Wigner cuts of the relevant states for the dual-rail code(6 cardinal states and the error state. The second row shows the 16 two-qubit Pauli expectation values extracted by measuring 16 points in phase space. Last row depicts the extracted logical Pauli expectation values of the dual-rail subspace. All the measured data shown includes SPAM errors. }
\label{fig_s5}
\end{figure}

We can easily extend this technique for our dual-rail qubit. Crucially, we will have to measure the joint-Wigner, defined as, 

\begin{equation}
    W(\alpha, \beta) = \frac{4}{\pi^2}\mathrm{Tr\big[ \rho\hat{P}_J(\alpha, \beta) \big]}
\end{equation}

where we define the displaced joint parity operator as $\hat{P}_J(\alpha, \beta) = \hat{P}_A(\alpha)\otimes\hat{P}_B(\beta)$. We first note that we cannot directly project on to the dual-rail subspace since, due to erasure errors, the system leaks out of this subspace. In order to incorporate the error state $\ket{00}$, we instead choose to restrict ourselves on to the two qubit subspace spanned by Fock qubit in both Alice and Bob : $\{\ket{0}, \ket{1}\}_A \otimes \{\ket{0}, \ket{1}\}_B$. The dual-rail subspace, including the error space, lives within this larger space, even though we have some states we are not concerned with. Now, projecting the displaced joint parity onto this subspace, we get, 

\begin{equation}
    (\hat{I}_A \otimes \hat{I}_B) \hat{P}_J(\alpha, \beta) (\hat{I}_A \otimes \hat{I}_B)  = (\hat{I}_A \hat{P}_A(\alpha) \hat{I}_A) \otimes (\hat{I}_B \hat{P}_B(\beta) \hat{I}_B)
\end{equation}

which we then equate to the 16 two-qubit Paulis $\{\hat{I}, \hat{X}, \hat{Y}, \hat{Z}\}_A \otimes \{\hat{I}, \hat{X}, \hat{Y}, \hat{Z}\}_B$. Its easy to see that this will yield 16 points, which are just combinations of the 4 points from single-mode displacements that we derived above. We write this set of 16 displacements explicitly below,

\begin{equation}
    \begin{split}
        d = (\alpha, \beta) = \,\, \frac{1}{\sqrt{2}} \times \bigg\{
        &(0, 0), (0, 1), (0, -1), (0, i),
        (1, 0), (1, 1), (1, -1), (1, i), \\
        &(-1, 0), (-1, 1), (-1, -1), (-1, i),
        (i, 0), (i, 1), (i, -1), (i, i), \bigg\}
    \end{split}
\end{equation}

We will denote $d_i$ as the $i$-th element in the above set, with $i = 0, 1, 2,..., 15$. From these displacements, we can construct the 16 two-qubit Pauli matrices as follows, 

\begin{equation}
    \begin{split}
        &\hat{I}_A\hat{I}_B = \,\frac{e^2}{4}\bigg[ \hat{P}_J(d_5) + \hat{P}_J(d_6) + \hat{P}_J(d_9) + \hat{P}_J(d_{10})\bigg]\\
        &\hat{I}_A\hat{X}_B = \,\frac{e^2}{4\sqrt{2}}\bigg[ \hat{P}_J(d_5) - \hat{P}_J(d_6) + \hat{P}_J(d_9) - \hat{P}_J(d_{10})\bigg]\\
        &\hat{I}_A\hat{Y}_B = \,\frac{e^2}{4\sqrt{2}}\bigg[ 2\hat{P}_J(d_7) - \hat{P}_J(d_5) - \hat{P}_J(d_6) + 2\hat{P}_J(d_{11}) - \hat{P}_J(d_{9}) -\hat{P}_J(d_{10})\bigg]\\
        &\hat{I}_A\hat{Z}_B = \,\frac{e}{2}\bigg[ \hat{P}_J(d_4) + \hat{P}_J(d_8)\bigg]\\
        &\hat{X}_A\hat{I}_B = \,\frac{e^2}{4\sqrt{2}}\bigg[ \hat{P}_J(d_5) + \hat{P}_J(d_6) - \hat{P}_J(d_9) - \hat{P}_J(d_{10})\bigg]\\
        &\hat{X}_A\hat{X}_B = \,\frac{e^2}{8}\bigg[ \hat{P}_J(d_5) - \hat{P}_J(d_6) - \hat{P}_J(d_9) + \hat{P}_J(d_{10})\bigg]\\
        &\hat{X}_A\hat{Y}_B = \,\frac{e^2}{8}\bigg[ 2\hat{P}_J(d_7) - \hat{P}_J(d_5) - \hat{P}_J(d_6) - 2\hat{P}_J(d_{11}) + \hat{P}_J(d_{9}) + \hat{P}_J(d_{10})\bigg]\\
        &\hat{X}_A\hat{Z}_B = \,\frac{e}{2\sqrt{2}}\bigg[ \hat{P}_J(d_4) - \hat{P}_J(d_8)\bigg]\\
        &\hat{Y}_A\hat{I}_B = \,\frac{e^2}{4\sqrt{2}}\bigg[ 2\hat{P}_J(d_{13}) + 2\hat{P}_J(d_{14}) - \hat{P}_J(d_5) - \hat{P}_J(d_{6}) - \hat{P}_J(d_{9}) -\hat{P}_J(d_{10})\bigg]\\
        &\hat{Y}_A\hat{X}_B = \,\frac{e^2}{8}\bigg[ 2\hat{P}_J(d_{13}) - 2\hat{P}_J(d_{14}) - \hat{P}_J(d_5) + \hat{P}_J(d_{6}) - \hat{P}_J(d_{9}) +\hat{P}_J(d_{10})\bigg]\\
        &\hat{Y}_A\hat{Y}_B = \,\frac{e^2}{8}\bigg[ 4\hat{P}_J(d_{15}) - 2\hat{P}_J(d_{13}) - 2\hat{P}_J(d_{14}) - 2\hat{P}_J(d_{7}) + \hat{P}_J(d_{5}) +\hat{P}_J(d_{6}) - 2\hat{P}_J(d_{11}) + \hat{P}_J(d_{9}) +\hat{P}_J(d_{10})\bigg] \\
        &\hat{Y}_A\hat{Z}_B = \,\frac{e^2}{2\sqrt{2}}\bigg[ \hat{P}_J(d_{12}) - \hat{P}_J(d_4) - \hat{P}_J(d_8)\bigg]\\
        &\hat{Z}_A\hat{I}_B = \,\frac{e}{2}\bigg[ \hat{P}_J(d_1) + \hat{P}_J(d_2)\bigg]\\
        &\hat{Z}_A\hat{X}_B = \,\frac{e}{2\sqrt{2}}\bigg[ \hat{P}_J(d_1) - \hat{P}_J(d_2) \bigg]\\
        &\hat{Z}_A\hat{Y}_B = \,\frac{e}{2\sqrt{2}}\bigg[ 2\hat{P}_J(d_3) - \hat{P}_J(d_1) - \hat{P}_J(d_2)\bigg]\\
        &\hat{Z}_A\hat{Z}_B = \, \hat{P}_J(d_0)\\
    \end{split}
\end{equation}

For our purposes, we only care about the Pauli operators for the dual-rail, which can be obtained by linear combinations of the above matrices, since we work in the $\{\ket{01}, \ket{10}\}$ basis. Therefore, the logical Pauli operators for the dual-rail qubit are

\begin{equation}
    \begin{split}
        \hat{I}_L &= \,\frac{1}{2}\bigg[ \hat{I}_A\hat{I}_B - \hat{Z}_A\hat{Z}_B\bigg]\\
        \hat{X}_L &= \,\frac{1}{2}\bigg[ \hat{X}_A\hat{X}_B + \hat{Y}_A\hat{Y}_B\bigg]\\
        \hat{Y}_L &= \,\frac{1}{2}\bigg[ \hat{Y}_A\hat{X}_B - \hat{X}_A\hat{Y}_B\bigg]\\
        \hat{Z}_L &= \,\frac{1}{2}\bigg[ \hat{Z}_A\hat{I}_B - \hat{I}_A\hat{Z}_B\bigg]\\
    \end{split}
\end{equation}

\cref{fig_s4}b shows the 16 points measured in phase space, depicted in different joint-Wigner cuts. \cref{fig_s4} shows the measurement of joint-Wigner cuts, the expectation values of the two-qubit Pauli operators and subsequently constructing the dual-rail Pauli expectation values from them. 

\subsection{Dual-rail Pauli transfer matrix with erasure detection}

\begin{figure}
\includegraphics{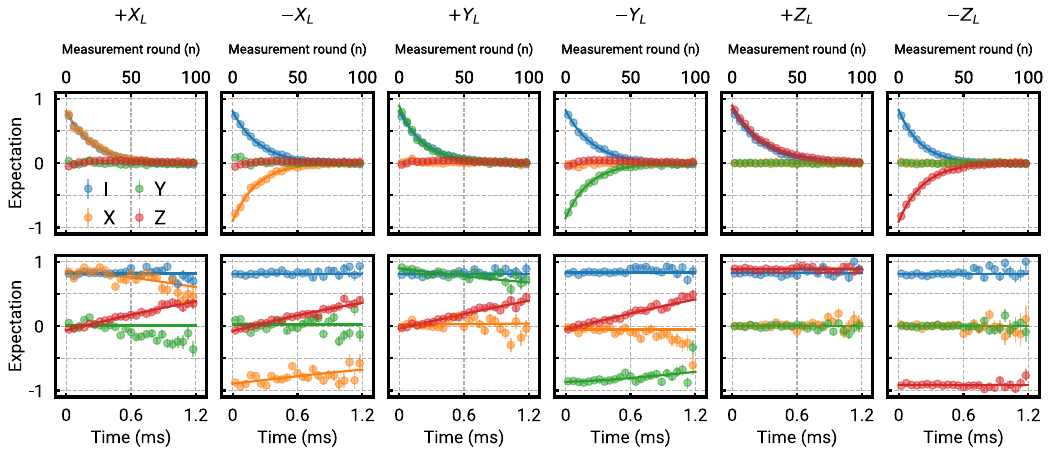}
\caption{\textbf{a,} Measured expectation values of the dual-rail Pauli operators as a function of number of erasure detection rounds, with solid lines shows the fits to exponential decay. \textbf{b,} Expectation values postselected on no jumps in the trajectory. $\langle \hat{I}(t) \rangle$ and $\langle \hat{Z}(t) \rangle$ are fit to exponential decay. Since these plots are extremely flat, we can only extract an upper bound on the residual leakage and bit-flip rate. $\langle \hat{X}(t) \rangle$ and $\langle \hat{Y}(t) \rangle$ are fit to the analytical no-jump evolution with an additional exponential decay term to extract residual dephasing rate.}
\label{fig_s6}
\end{figure}

We measure the expectation values of the Pauli operators after preparing the 6 cardinal states as a function of number of erasure detection rounds. Top panel in \cref{fig_s6} shows the measured values with exponential fits. The decay of the identity operator clearly shows leakage outside of the codespace. The bottom panel shows postselected data. The postselected $\langle \hat{I} \rangle$ looks constant and we can only obtain an upper bound on residual leakage rate by fitting to an exponential decay $(<10^{-4} (\mathrm{ms})^{-1})$. We then fit the postselected $\langle \hat{X} \rangle$ and $\langle \hat{Y} \rangle$ to extract dephasing rates within the dualrail subspace. Taking into account the no-jump evolution, the analytical formula for the evolution of these operators will look like \cite{teoh_dual-rail_2023}

\begin{equation}
    \langle \hat{X}(t) \rangle = \frac{(W^*V+V^*W)e^{-\frac{1}{2}\Delta\kappa t}}{|V|^2 + |W|^2e^{-\Delta\kappa t}}
\end{equation}

\begin{equation}
    \langle \hat{Y}(t) \rangle = \frac{i(W^*V-V^*W)e^{-\frac{1}{2}\Delta\kappa t}}{|V|^2 + |W|^2e^{-\Delta\kappa t}}
\end{equation}

\begin{equation}
    \langle \hat{Z}(t) \rangle = \frac{(|V|^2-|W|^2)e^{-\frac{1}{2}\Delta\kappa t}}{|V|^2 + |W|^2e^{-\Delta\kappa t}}
\end{equation}

\noindent
for an arbitrary state $\ket{\psi} = V\ket{01}+W\ket{10}$. On top of this, we add an exponentially decaying term to calculate residual dephasing and bit-flip rate within the codespace. We observe that the fits agree well with the measured data. We note that the increase in $\langle \hat{Z} \rangle$ for the equator states shows the polarisation of the Bloch vector towards the longer lived cavity (Bob) due to the no-jump backaction.